\documentclass[11pt]{article}
\usepackage{amsfonts,amssymb, graphicx,bm}
\usepackage{psfrag}
\newcommand{\ur}{\mathbf{r}}
\newcommand{\us}{\mathbf{s}}

\newcommand{\N}{\mathbb{N}}
\newcommand{\R}{\mathbb{R}}
\newcommand{\C}{\mathbb{C}}
\newcommand{\F}{\mathsf{F}}
\newcommand{\T}{\mathsf{T}}
\newcommand{\PP}{P_\ast}
\newcommand{\ga}{\gamma}
\newcommand{\PO}{\stackrel{\circ}{P}}
\def\bbone{{\mathchoice {\rm 1\mskip-4mu l} {\rm 1\mskip-4mu l}
{\rm 1\mskip-4.5mu l} {\rm 1\mskip-5mu l}}}
\newcommand{\qed}{\hfill\rule{3mm}{3mm}}
\newcommand{\Bsp}{\qquad\qquad\qquad\qquad}
\newtheorem{Lemma}{Lemma}

\newtheorem{Theorem}{Theorem}

\begin{document}
\title{The Ground State Energy of The Massless Spin-Boson Model}
\author{Abdelmalek Abdesselam}
\maketitle

\begin{center}
{\it Department of Mathematics,
P. O. Box 400137,
University of Virginia,
Charlottesville, VA 22904-4137, USA }\\
email: \texttt{malek@virginia.edu}
\end{center}

\bigskip

\parbox{11.8cm}{ \small
{\bf Abstract.}
We provide an explicit combinatorial expansion for the ground state energy
of the massless spin-Boson model as a power series in the coupling parameter.
Our method uses the technique of cluster expansion in constructive quantum field theory
and takes as a starting point the functional integral representation and its reduction
to an Ising model on the real line with long range interactions.
We prove the analyticity of our expansion and provide an explicit lower bound
on the radius of convergence.
We do not need multiscale nor renormalization group analysis.
A connection to the loop-erased random walk is indicated.
}

\bigskip

\parbox{12cm}{\small
}

\tableofcontents

\section{Introduction}

The spin-Boson model is one of the simplest quantum mechanical models for the interaction of matter with
radiation. Yet, many of its properties, especially scattering in the massless case, remain
a mathematical mystery.
It is (formally) given by the Hamiltonian
\[
H_\lambda=(\sigma_z+I)\otimes I+
I\otimes\int_{\R^3} \omega(k)\ a^\ast(k) a(k)\ d^3k
\]
\[
+\lambda\ \sigma_x\otimes\ \int_{\R^3} 
\frac{f(k) a^\ast (k)+\overline{f(k)} a(k)}{4\pi\sqrt{\omega(k)}}
\ d^3k\ .
\]

The Hilbert space is $\C^2\otimes \mathcal{F}$ where $\mathcal{F}$ is the Bosonic
Fock space built on $\mathfrak{h}=L^2(\R^3,\C,d^3k)$.
The creation and annihilation operators, $a^\ast(k)$ and $a(k)$ respectively, 
satisfy the usual canonical commutation relations.
The matrices $\sigma_x$ and $\sigma_z$ are standard Pauli matrices.
We refer to~\cite{HaslerH}
for a more detailed spectral theoretic presentation of the model.
For ease of reference, we kept the exact same formulation and notations as in the cited article.
For more background from a physical perspective,
the reader may consult~\cite{LeggettCDFGZ} for a good review
of the spin-Boson model as well as its Fermionic cousin: the Kondo model.
For a review from a mathematical perspective, we recommend~\cite{HubnerS}, where many open problems
are listed.

It is well-known that if $\frac{f}{\sqrt{\omega}}$ and $\frac{f}{\omega}$ belong to $\mathfrak{h}$,
and for real $\lambda$, then $H_{\lambda}$ is self-adjoint and bounded below.
Of particular interest is the massless case where $\omega(k)=|k|$ and $f$ does not vanish at the origin.
Because of the infrared singularity at $k=0$, the mathematical analysis in that situation is more difficult.
Once self-adjointness is established, the first step towards the understanding of the dynamics of the model is to construct the ground state $\Omega_{GS}(\lambda)$ and its energy
$E(\lambda)=\inf \sigma(H_\lambda)$.
As this energy is at the bottom of the continuous spectrum when $\lambda=0$, the standard perturbation
theory tools (as one can find say in~\cite{Kato}) do not apply.
More sophisticated methods are needed for the control of the ground state energy
for nonzero coupling $\lambda$.
Such a method was introduced by Bach, Fr\"ohlich and Sigal in their seminal work~\cite{BachFS1,BachFS2} and it has been further 
developed over the last decade (see, e.g.,~\cite{BachCFS,BachFP1,BachFP2}).
One of its main motivations is to provide rigorous constructive algorithms for the calculation of quantities such as $E(\lambda)$, in the case of massless Bosons.
Such an algorithm based on the renormalization group (RG) in operator space, via the Feshbach
transformation, is in principle powerful enough to show analyticity results for $E(\lambda)$.
The full delivery on that promise, for the spin-Boson model, has been made only very recently 
in~\cite{HaslerH} where the construction of $E(\lambda)$
and $\Omega_{GS}(\lambda)$ and the proof of their analyticity was obtained.
In fact, two constructive algorithms are given in~\cite{HaslerH}.
The first one is to follow the RG iterations.
The second, which is computationally much simpler, can be summarized by saying:
put an infrared cut-off, compute the Rayleigh-Schr\"odinger (RS) pertubation series then remove
the cut-off. At the end of this process, all the terms of the series are free of infrared divergences.
Moreover, the series converges for small $\lambda$.
Unfortunately, the first algorithm is needed to prove these statements, which are very far
from obvious
when looking at the RS series.
Indeed, by looking at the first few low orders in perturbation, Hasler and Herbst found
some rather mysterious cancellations of infrared divergences.
By their RG proof, one knows in an indirect way that these cancellations must persist
through all orders in $\lambda$. Prior to~\cite{HaslerH}, analyticity of the ground state
energy with respect to $\lambda$ (and other parameters) was established
for less singular nonrelativistic QED
models, using the RG techniques of Bach, Fr\"ohlich and Sigal, in the article~\cite{GriesemerH}.
On Page 580 of the latter, one can read: ``It seems unlikely that another approach, not based on a
renormalization analysis would yield a result similar to ours''.
As demonstrated in the present article, {\it there is another way} !

Another expression for $E(\lambda)$ was found in~\cite{Hirokawa} but it is far from explicit,
nor is it clear (at least to us) how one can use it to prove analyticity.
In this article, we obtain a completely explicit combinatorial expansion for $E(\lambda)$.
It is explicit to the point one can see that it is analytic {\it literally} just by looking
at it! By ``it'' we mean the graphs produced by our expansion as in the figure at the end
of \S 4 below.
Moreover, the new method we introduce in this article does not involve multiscale analysis nor
renormalization, despite the massless situation.
Our method is a direct descendent of the
infamous combination of the Euclidean strategy {\it plus} the cluster
expansion technology which worked so well in the context of constructive quantum field theory (CQFT).
We are not the first to use this general approach in the context of nonrelativistic QED, see,
e.g.,~\cite{LorincziM}. However, we have the advantage of using the latest `stable version'
of the CQFT cluster expansion software~\cite{Abdesselam1} which comes with such optional features
as multiscale analysis and renormalization, large versus small field analysis, $p$-particle irreducibility analysis, and could also handle microlocal sectorial decomposition as
in~\cite{Poirot,DisertoriR}. Since we do not need these in this article, we keep them in reserve
for more dire infrared (or ultraviolet) future circumstances.
There is also a very promising new set of ideas on how to improve CQFT expansion techniques
(see~\cite{MagnenR,GurauMR}), but this is still at the `beta version' stage.
These ideas which originated in~\cite{RivasseauJHEP} unexpectedly came from the study of renormalization in noncommutative
quantum field theory (see, e.g.,~\cite{RivasseauV}).   
A good pedagogical entry point in the subject of CQFT expansions
is~\cite{Rivasseau} (see also the follow up research article~\cite{RivasseauW}).

Our point of departure is Bloch's formula
\[
E(\lambda)=\lim_{T\rightarrow \infty}\ -\frac{1}{T}\log
\left(\Omega_{\downarrow}, e^{-T H_{\lambda}} \Omega_{\downarrow}\right)
\]
as well as the Feynman-Kac-Nelson formula followed by integration over the Bosons, with the result
\begin{equation}
\left(\Omega_{\downarrow}, e^{-T H_{\lambda}} \Omega_{\downarrow}\right)
=Z\left(\left(\frac{\lambda}{4\pi}\right)^2,\ T\right)
\label{reduc}
\end{equation}
where
\begin{equation}
Z(\alpha, T)=\mathbb{E}\left[
\exp\left(
\frac{\alpha}{2}
\int_0^T\int_0^T\ X(t)X(s)h(t-s)\ dt\ ds
\right)
\right]\ .
\label{Zdef}
\end{equation}
The vector $\Omega_{\downarrow}$
is the free vacuum.
The function $h(s)$, $s\in\R$, is defined by
\[
h(s)=\int_{\R^3} \frac{|f(k)|^2}{\omega(k)}\ e^{-|s|\omega(k)}\ d^3k\ge 0\ .
\]
Under the hypotheses on $f$, it is continuous and satisfies $||h||_{L^\infty}=
||\frac{f}{\sqrt{\omega}}||_{L^2}$.
The finite time (or volume) partition function $Z(\alpha,T)$
is an expectation over a continuous time Markov jump process $X(t)$, $t\ge 0$, taking values $\pm 1$.
The boundary condition at $t=0$ is free: $\mathbb{P}(X(0)=1)=\mathbb{P}(X(0)=-1)=\frac{1}{2}$.
The wait times between two consecutive spin flips are exponentially distributed with unit parameter
(i.e., with measure $e^{-t}dt$).
This is the famous Ising model over $\R$ with long range $\sim\frac{1}{t^2}$ 
interactions introduced in~\cite{AndersonY} for the Kondo model and in~\cite{EmeryL} for spin-Boson
model. 
See~\cite{SpohnD,FannesN,Spohn} for a mathematical study of the spin-Boson model via this route.
Our main result is the following.

\medskip
\noindent{\bf Claim:}
{\it One has an explicit power series expansion for
\[
\lim_{T\rightarrow\infty}\frac{-1}{T}\log Z(\alpha,T)\ . 
\]
This quantity is analytic, with a lower bound on the radius of convergence
in $\alpha$ given by
\[
R_{\rm min}=\left[
32\times \sqrt{e}\times \max\left(||h||_{L^\infty},||h||_{L^1}\right)
\right]^{-1}\ .
\]
}

\medskip
We defer the precise statement of the theorem to \S4 since a fair amount of combinatorics must be introduced beforehand.
Note that we do not yet have a lower bound on the radius in $\lambda$ for the ground state energy.
This is because Bloch's formula requires a
nonzero overlap $(\Omega_{\downarrow},\Omega_{GS}(\lambda))\neq 0$ which is known to hold for
small coupling, for instance by the results of~\cite{HaslerH}, but with no explicit range of
validity.
This will be fixed in~\cite{Abdesselam5} where the present expansion technique is used
for the construction of the ground state itself.

We will not dwell on the the proof of the reduction (\ref{reduc})
which is standard. One needs to use the Schr\"odinger representation of $\mathcal{F}$
in $Q$-space (see, e.g.,~\cite{GlimmJ,Simon}) as well as the Feynman-Kac-Nelson formula
(see, e.g.,~\cite{Simon,KleinL,Hiroshima} and also~\cite{HiroshimaL}
for the simultaneous treatment of spin).
Note that the spin part of the interaction can be made to act as a multiplication operator.
This is done by a simple conjugation with $u=\frac{1}{\sqrt{2}}\left(
\begin{array}{cc}
1 & -1\\
1 & 1
\end{array}
\right)$, namely, using $u^{-1}(\sigma_z+I)u=I-\sigma_x$
for the free Hamilatonian and $u^{-1} \sigma_x u=\sigma_z$ for the interaction.
The jump process is the one corresponding to the semi-group $e^{-t(I-\sigma_x)}$.
Finally the vector $u^{-1}\left(
\begin{array}{c}
0\\
1
\end{array}
\right)$
coming from the spin-down choice $\Omega_{\downarrow}$, is responsible for the boundary condition
on $X(0)$.

By contrast, the combinatorics of our expansion will be explained in great
detail.
When used in problems of mathematical analysis, combinatorics is a bit like a Toruk~\cite{Avatar}
--- it can be a powerful friend, but it must be approached with respect.
The degree of detail serves the purpose of showing, by example, what we mean by
respect for combinatorics.
Over and over in the course of our proof, we will use relabeling changes of variables, continuous or discrete. This is because behind the scene there is Joyal's theory of combinatorial
species~\cite{Joyal,BergeronLL} at
work. Although we do not resort to this theory explicitly, in this article, 
it is the proper mathematical framework for much of the combinatorics involved
in Feynman/cluster type expansions.
The two references mentioned only consider applications to the field of
combinatorial enumeration. The weights used are placeholders such as
$x^n$ where $n$ counts something. In accordance with Sokal's multivariate
philosophy~\cite{Sokal},
allowing more general weights such as {\it contractions of tensors} is where the full potential
of the theory is. Some steps in this direction were taken in~\cite{Abdesselam2}. From the point
of view of the latter article, the theory of species becomes a {\it lingua franca} shared by such remote
areas as quantum field theory and (neo)classical invariant theory. The only difference
is that of infinite versus finite dimension respectively. More on this can be found in~\cite{AbdesselamB,AbdesselamC,Abdesselam3}.

This article demonstrates that there is a bright future
for `the Euclidean strategy $+$ CQFT cluster expansion' method in the area of
nonrelativistic QED. The first part, i.e., the Euclidean reformulation
of the problems posed by the interaction of matter with radiation
should be straightforward.
To paraphrase the prophetic words of Paul Federbush~\cite{Federbush},
the second part is where the action is.
One can try to use techniques similar to ours, in order to treat progressively
more difficult and more realistic models.
However, we think it is more important, at this point in time, to focus
on the simplest models but try to prove more, i.e., to try to understand scattering completely.
As advocated in~\cite{HubnerS}, the massless spin-Boson model is definitely in this class.
There is another one which is even simpler: the Nelson model with quadratic confining potential.
We take this opportunity to announce the obtention of the exact analogue of the result in the present
article, for this particular case of Nelson's model (above three dimensions)~\cite{AbdesselamH}.
We are encouraged by how forgiving the combinatorics of these two models are,
and therefore led to believe
that a deeper understanding of scattering is within reach.
The ultimate goal, if this endevor succeeds, would be to apply the lessons learned
in such simpler models to the scattering theory of $P(\phi)_2$.
The latter is the missing piece of the glorious CQFT work in the seventies (see~\cite{AuilB}
for a recent update and list of references).
We would like to dream that this achievement is possible.

\section{The jump process}

Let $N(t)$, $t\ge 0$, be the usual Poisson process with parameter $1$.
Let $B=\pm 1$ be an independent Bernoulli random variable with $\mathbb{P}(B=1)=\mathbb{P}(B=-1)=\frac{1}{2}$.
We realize the jump process $X(t)$, $t\ge 0$ as
\[
X(t)=B\cdot (-1)^{N(t)}\ .
\]

Clearly, $Z(\alpha,T)$ defined in (\ref{Zdef}) is an
entire function of $\alpha\in\C$,
for any given finite time $T$.
Indeed, by the Fubini-Tonelli Theorem one has the everywhere convergent power series expansion
\[
Z(\alpha,T)=\sum_{p=0}^{\infty}
\left(\frac{\alpha}{2}\right)^p\frac{1}{p!}
\int_{[0,T]^{2p}}
\prod_{i=1}^{2p} dt_i\Bsp
\]
\[
\Bsp
\prod_{j=1}^{p} h(t_{2j}-t_{2j-1})
\ \times\ \mathbb{E}\left[
X(t_1)\cdots X(t_{2p})
\right]\ .
\]
Our first task is to explicitly evaluate the expectation.
\begin{Lemma}
Let $0<t_1<\cdots<t_q$ be an increasing sequence of times, then
\[
\mathbb{E}\left[
X(t_1)\cdots X(t_q)
\right]=\left\{
\begin{array}{cc}
e^{-2[|t_1-t_2|+|t_3-t_4|+\cdots+|t_{q-1}-t_{q}|]} & {\rm if}\ q\ {\rm is\ even},\\
0 & {\rm if}\ q\ {\rm is\ odd}.
\end{array}
\right.
\]
\end{Lemma}

\noindent{\bf Proof:}
Write
\[
X(t_1)\cdots X(t_q)=B^q\cdot (-1)^{N(t_1)+\cdots+N(t_q)}\ .
\]
Then, by Abel's summation by parts
\[
N(t_1)+\cdots+N(t_q)=q N(t_1)+(q-1)(N(t_2)-N(t_1))+\cdots+(N(t_q)-N(t_{q-1}))\ .
\]
By the independence of the increments
\[
\mathbb{E}\left[
X(t_1)\cdots X(t_q)
\right]=\mathbb{E}[B^q]\ \mathbb{E}\left[
(-1)^{qN(t_1)}
\right]\Bsp
\]
\[
\Bsp
\times\mathbb{E}\left[
(-1)^{(q-1)(N(t_2)-N(t_1))}
\right]\cdots
\mathbb{E}\left[
(-1)^{N(t_q)-N(t_{q-1})}
\right]\ .
\]
This vanishes if $q$ is odd because $\mathbb{E}(B^q)=\mathbb{E}(B)=0$.
Now assume $q$ is even, then
\[
\mathbb{E}\left[
X(t_1)\cdots X(t_q)
\right]=
\mathbb{E}\left[
(-1)^{N(t_2)-N(t_1)}
\right]\Bsp
\]
\[
\Bsp
\times\mathbb{E}\left[
(-1)^{N(t_4)-N(t_3)}
\right]
\cdots
\mathbb{E}\left[
(-1)^{N(t_q)-N(t_{q-1})}
\right]\ .
\]
Since by the definition of $N$, $N(t_i)-N(t_{i-1})$ is distributed
according to the Poisson distribution with parameter $t_{i}-t_{i-1}$,
\[
\mathbb{E}\left[
(-1)^{N(t_i)-N(t_{i-1})}
\right]=
\sum_{k=0}^{\infty} (-1)^{k} e^{-(t_i-t_{i-1})} \frac{(t_{i}-t_{i-1})^k}{k!}= e^{-2(t_i-t_{i-1})}
\]
and the lemma follows.
\qed

\section{Preparation for the Mayer/cluster expansion}

In order to use the previous lemma, we need to break the cube of integration over times into
$(2p)!$ simplices.
We therefore write
\begin{equation}
Z(\alpha,T)=\sum_{p=0}^{\infty}
\left(\frac{\alpha}{2}\right)^p\frac{1}{p!}
\sum_{\sigma\in\mathfrak{S}_{2p}}
\int_{0<t_{\sigma(1)}<\cdots<t_{\sigma(2p)}<T}
\prod_{i=1}^{2p} dt_i\Bsp
\]
\[
\Bsp
\prod_{j=1}^{p} h(t_{2j}-t_{2j-1})
\times\prod_{j=1}^{p} e^{-2|t_{\sigma(2j)}-t_{\sigma(2j-1)}|}\ .
\end{equation}
We find it more convenient to work
with a fixed simplex, so we perform a change of variables
\[
(t_{\sigma(1)},\ldots,t_{\sigma(2p)})\rightarrow(t_1,\ldots,t_{2p})
\]
for given fixed permutation $\sigma$.
We get
\[
Z(\alpha,T)=\sum_{p=0}^{\infty}
\left(\frac{\alpha}{2}\right)^p\frac{1}{p!}
\sum_{\sigma\in\mathfrak{S}_{2p}}
\int_{0<t_{1}<\cdots<t_{2p}<T}
\prod_{i=1}^{2p} dt_i\Bsp
\]
\[
\Bsp
\prod_{j=1}^{p} h(t_{\sigma^{-1}(2j)}-t_{\sigma^{-1}(2j-1)})
\times\prod_{j=1}^{p} e^{-2|t_{2j}-t_{2j-1}|}\ .
\]
We will also change variables $\sigma^{-1}\rightarrow\sigma$ in the sum over permutations in order to lighten the notations, i.e.,
have $t_{\sigma(2j)}-t_{\sigma(2j-1)}$
instead of $t_{\sigma^{-1}(2j)}-t_{\sigma^{-1}(2j-1)}$ as arguments of the $h$ functions.

Now let us recall some elementary notions of graph theory.
We will use the notation $[n]$ for the set of integers $\{1,2,\ldots,n\}$.
When considering a simple graph $G=(V,E)$ with fixed vertex set $V$,
we will identify $G$ with its edge set $E$ seen as a set of unordered pairs in $V$.
Namely, let $\mathcal{P}(V)$ denote the power set of $V$, and let
$V^{(2)}=\{A\in \mathcal{P}(V)|\ \ |A|=2\}$ where $|\cdot|$ denotes cardinality.
We will view $E$ (as well as $G$) as a subset of $V^{(2)}$.
If $G$ is a multigraph with (unlabeled) repeated edges but no loops (`tadpoles' for physicists),
we will instead view $G\simeq E$ as a multiset of edges, i.e.,
an element of $\N^{V^{(2)}}$.
The degree of a vertex is the number of edges incident to $v$ and it is denoted
by ${\rm deg}_{G}(v)$.
A $k$-factor on $V$ is a graph $G$ such that ${\rm deg}_{G}(v)=k$
for all $v\in V$.
A perfect matching is a $1$-factor.

Now let us rewrite $Z(\alpha,T)$ in terms of perfect matchings.
For $p\ge 0$ and $\sigma\in\mathfrak{S}_{2p}$, let $P(\sigma)$
be the perfect matching on $[2p]$ given by
\[
P(\sigma)=\left\{
\{\sigma(1),\sigma(2)\},\{\sigma(3),\sigma(4)\},\ldots,\{\sigma(2p-1),\sigma(2p)\}
\right\}\ .
\]
The one corresponding to the identity permutation $\sigma=Id$ will be denoted by
$\PP$ which is an analogue of a base point.
Now by collecting permutations $\sigma$ which give the same perfect matching $P$ we have
\[
Z(\alpha,T)=\sum_{p=0}^{\infty} \alpha^p
\sum_{P}
\int_{0<t_{1}<\cdots<t_{2p}<T}
\prod_{i=1}^{2p} dt_i\Bsp
\]
\[
\Bsp
\prod_{\{a,b\}\in P} h(t_a-t_b)
\times
\prod_{\{a,b\}\in\PP} e^{-2|t_a-t_b|}
\]
where the second sum is over all the $\frac{(2p)!}{2^p p!}$ perfect matchings on $[2p]$.
Indeed, for any $P$ there are $2^p p!$ permutations $\sigma$ for which $P(\sigma)=P$.

Now we will partially undo what we just did.
By writing the number 1 in a more complicated way, we have
\[
Z(\alpha,T)=\sum_{p=0}^{\infty} \alpha^p
\sum_{P}
\left(\frac{1}{p!}\sum_{\tau\in\mathfrak{S}(\PP)} 1\right)
\int_{0<t_{1}<\cdots<t_{2p}<T}
\prod_{i=1}^{2p} dt_i\Bsp
\]
\[
\Bsp
\prod_{\{a,b\}\in P} h(t_a-t_b)
\times
\prod_{\{a,b\}\in\PP} e^{-2|t_a-t_b|}
\]
where $\mathfrak{S}(\PP)$
denotes the set of permutations of the set $\PP$.
Now by Fubini's Theorem
\[
Z(\alpha,T)=\sum_{p=0}^{\infty} \frac{\alpha^p}{p!}
\sum_{\tau\in\mathfrak{S}(\PP)}
\int_{0<t_{1}<\cdots<t_{2p}<T}
\prod_{i=1}^{2p} dt_i\Bsp
\]
\[
\Bsp
\left\{
\sum_{P}
\prod_{\{a,b\}\in P} h(t_a-t_b)
\right\}
\times
\prod_{\{a,b\}\in\PP} e^{-2|t_a-t_b|}\ .
\]
Given $\tau\in\mathfrak{S}(\PP)$ define an induced permutation
$\sigma[\tau]\in\mathfrak([2p])$ by
\[
\left\{
\begin{array}{ccc}
\sigma[\tau](2i-1) & = & \min \tau\left(\{2i-1,2i\}\right)\\
\sigma[\tau](2i) & = & \max \tau\left(\{2i-1,2i\}\right)
\end{array}
\right.
\]
i.e., this is the permutation on the underlying set of points $[2p]$
obtained by rigidly permuting the pair-blocks of $\PP$ according to $\tau$.
Clearly, $\PP$, seen as a set partition of $[2p]$,
is invariant by $\sigma[\tau]$ so
\[
\prod_{\{a,b\}\in\PP} e^{-2|t_a-t_b|}
=
\prod_{\{a,b\}\in\PP} e^{-2|t_{\sigma[\tau]^{-1}(a)}-t_{\sigma[\tau]^{-1}(b)}|}\ .
\]
Besides,
\[
\sum_{P}
\prod_{\{a,b\}\in P} h(t_a-t_b)
=
\sum_{P}
\prod_{\{a,b\}\in P} h(t_{\sigma[\tau]^{-1}(a)}-t_{\sigma[\tau]^{-1}(b)})
\]
since one can sum over the perfect matching
$P^{\tau}=\{\sigma[\tau](A)| A\in P\}$ instead of $P$.
This is just another (discrete) change of variables $P\rightarrow P^{\tau}$ due to relabeling.
Now do yet another change of variables
\[
(t_{\sigma[\tau]^{-1}(1)},\ldots,t_{\sigma[\tau]^{-1}(2p)})
\rightarrow
(t_1,\ldots,t_{2p})
\]
which results in the equality
\[
\int_{0<t_{1}<\cdots<t_{2p}<T}
\prod_{i=1}^{2p} dt_i\ 
\left\{
\sum_{P}
\prod_{\{a,b\}\in P} h(t_a-t_b)
\right\}
\times
\prod_{\{a,b\}\in\PP} e^{-2|t_a-t_b|}
\]
\[
=
\int_{0<t_{\sigma[\tau](1)}<\cdots<t_{\sigma[\tau](2p)}<T}
\prod_{i=1}^{2p} dt_i\ 
\left\{
\sum_{P}
\prod_{\{a,b\}\in P} h(t_a-t_b)
\right\}
\times
\prod_{\{a,b\}\in\PP} e^{-2|t_a-t_b|}\ .
\]
This way we keep the integrand the same for all $\tau$'s.
The sum over these permutations simply rebuilds a larger domain of integration.
Namely,
\[
Z(\alpha,T)=\sum_{p=0}^{\infty} \frac{\alpha^p}{p!}
\int_{[0,T]^{2p}}
\prod_{i=1}^{2p} dt_i\ 
\ 
\prod_{j=1}^{p}
\bbone\{t_{2j-1}<t_{2j}\}
\]
\[
\times
\bbone\left\{
\begin{array}{c}
{\rm the\ intervals}\\
{[t_{2j-1},t_{2j}]}\\
{\rm are\ disjoint}
\end{array}
\right\}
\times
\left\{
\sum_{P}
\prod_{\{a,b\}\in P} h(t_a-t_b)
\right\}
\times
\prod_{\{a,b\}\in\PP} e^{-2|t_a-t_b|}
\]
where $\bbone\{\cdots\}$ means the sharp characteristic function of the condition between braces.
As a result
\[
Z(\alpha,T)=\sum_{p=0}^{\infty} \frac{\alpha^p}{p!}
\sum_{P}
\int_{[0,T]^{2p}}
\prod_{i=1}^{2p} dt_i
\ 
\prod_{j=1}^{p}
\bbone\{t_{2j-1}<t_{2j}\}
\]
\[
\times
\prod_{\{a,b\}\in P} h(t_a-t_b)
\times
\prod_{\{a,b\}\in\PP} e^{-2|t_a-t_b|}
\]
\begin{equation}
\times
\prod_{1\le i<j\le p}
\bbone\left\{
[t_{2i-1},t_{2i}]\cap[t_{2j-1},t_{2j}]=
\emptyset\right\}
\label{ready}
\end{equation}
i.e., we are ready for a Mayer/cluster expansion.

\section{The expansion}

Firstly, given a perfect matching $P$ and the fixed matching
$\PP$, build the $2$-factor $\PP+P$ obtained by superimposing
the edges of $P$ with those of $\PP$.
Note that $\PP+P$ can be a multigraph
since some edges can become doubled.
Clearly, $\PP+P$ breaks up as a disjoint collection
of cycles of even length.
An example is
\[
\parbox{6cm}{\psfrag{1}{$\scriptstyle{1}$}
\psfrag{1}{$\scriptstyle{1}$}
\psfrag{2}{$\scriptstyle{2}$}
\psfrag{3}{$\scriptstyle{3}$}
\psfrag{4}{$\scriptstyle{4}$}
\psfrag{5}{$\scriptstyle{5}$}
\psfrag{6}{$\scriptstyle{6}$}
\psfrag{7}{$\scriptstyle{7}$}
\psfrag{8}{$\scriptstyle{8}$}
\psfrag{9}{$\scriptstyle{9}$}
\psfrag{10}{$\scriptstyle{10}$}
\psfrag{11}{$\scriptstyle{11}$}
\psfrag{12}{$\scriptstyle{12}$}
\psfrag{13}{$\scriptstyle{13}$}
\psfrag{14}{$\scriptstyle{14}$}
\psfrag{15}{$\scriptstyle{15}$}
\psfrag{16}{$\scriptstyle{16}$}
\psfrag{17}{$\scriptstyle{17}$}
\psfrag{18}{$\scriptstyle{18}$}
\psfrag{19}{$\scriptstyle{19}$}
\psfrag{20}{$\scriptstyle{20}$}
\includegraphics[width=6cm]{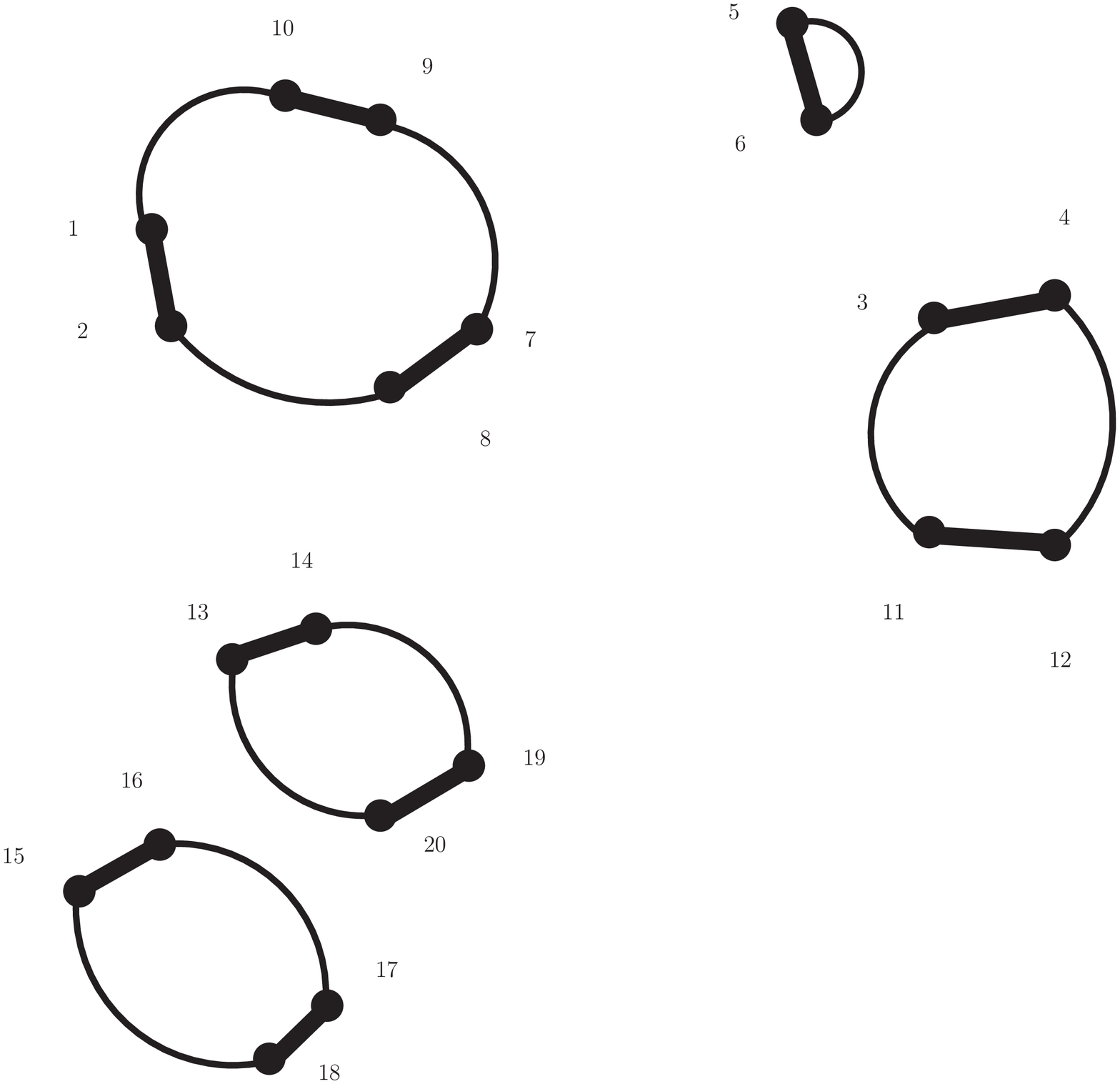}}
\]
where we represented the edges of $\PP$ by thick lines and those of $P$ by thinner ones.
Let $\pi(\PP+P)$ be the partition of the vertex set
$[2p]$ made of connected components of $\PP+P$, i.e., the collection of supports of these cycles.
Since $P$ and $\PP$ also are set partitions, another way of seeing
$\pi(\PP+P)$
is a $\PP\vee P$, the join in the partition lattice of $[2p]$
where the ordering $\preceq$ means ``is finer than or equal to''.
We also denote by $\Pi(\PP+P)$
the partition of $\PP$
induced by the partition $\pi(\PP+P)$ of $[2p]$.
Namely,
\[
\Pi(\PP+P)=\left\{
X\in\mathcal{P}(\PP)|
\ \cup_{A\in X} A\in \pi(\PP+P)
\right\}\ .
\]
This is the same as the connected component partition of the graph on $\PP$
obtained from $\PP+P$
by contracting the pairs of $\PP$ into points (and removing loops if any).
For a pair $A=\{2i-1,2i\}\in\PP$
we will use the notation $t_A$
for the interval $[t_{2i-1},t_{2i}]$.

Now write
\[
\prod_{1\le i<j\le p}
\bbone\left\{
[t_{2i-1},t_{2i}]\cap[t_{2j-1},t_{2j}]=\emptyset
\right\}=
\prod_{\{A,B\}\in\PP^{(2)}} \bbone\left\{t_A\cap t_B=\emptyset\right\}
\]
\[
=F(\mathbf{1})
\]
where the function
\[
\begin{array}{cl}
F: & \R^{\Pi(\PP+P)^{(2)}} \longrightarrow \R \\
{\ } & \us=\left(s_{\{X,Y\}}\right)_{\{X,Y\}\in \Pi(\PP+P)^{(2)}}  \longmapsto F(\us)
\end{array}
\]
is defined as follows, and $\mathbf{1}$ is the parameter vector
with all components equal to $1$.
By definition, we set
\[
F(\us)=
\prod_{X\in\Pi(\PP+P)}\left(\prod_{\{A,B\}\in X^{(2)}}
\bbone\left\{t_A\cap t_B=\emptyset\right\}\right)
\]
\[
\times
\prod_{\{X,Y\}\in\Pi(\PP+P)^{(2)}}
\left[\prod_{A\in X, B\in Y} \left(1-s_{\{X,Y\}}\bbone\left\{t_A\cap t_B\neq\emptyset\right\}
\right)\right]\ .
\]

Now use the Brydges-Kennedy-Abdesselam-Rivasseau formula~\cite{BrydgesK,AbdesselamR}
(see~\cite{Abdesselam4} for a gentle introduction and detailed proof of this identity).
The outcome is
\[
F(\mathbf{1})=\sum_{{\mathfrak{F}\ {\rm forest}}\atop{{\rm on}\ \Pi(\PP+P)}}
\int_{[0,1]^{\mathfrak{F}}}
\prod_{l\in\mathfrak{F}} du_l\ 
\frac{\partial^{|\mathfrak{F}|} F}{\prod_{l\in\mathfrak{F}} \partial s_l}
\left(\us(\mathfrak{F},u)\right)\ .
\]
Here the sum is over all forests $\mathfrak{F}$
on the vertex set $\Pi(\PP+P)$, i.e., simple graphs with no
cycles.
For each edge $l\in\mathfrak{F}$
one associates a real variable $u_l$ integrated between $0$ and $1$.
The integrand is the partial derivative of $F$ with respect to
the entries $s_{\{X,Y\}}$
such that $\{X,Y\}\in\mathfrak{F}$, evaluated
at the vector $\us(\mathfrak{F},u)=
\left(s(\mathfrak{F},u)_{\{X,Y\}}\right)_{\{X,Y\}\in\Pi(\PP+P)^{(2)}}$
defined by the following rule.
\begin{itemize}
\item
If $X$ and $Y$ belong to different connected components of $\mathfrak{F}$,
then $s(\mathfrak{F},u)_{\{X,Y\}}=0$.
\item
Otherwise, if $X$ and $Y$ are in the same (tree) component
of $\mathfrak{F}$, then  $s(\mathfrak{F},u)_{\{X,Y\}}=\min_{l} u_l$
where $l$ ranges over the unique simple path joining $X$ and $Y$,
in the forest $\mathfrak{F}$.
\end{itemize}

The outcome of the formula involves, for each $\{X,Y\}\in\mathfrak{F}$,
derivatives of the form
\[
\frac{\partial}{\partial s_{\{X,Y\}}}
\prod_{A\in X, B\in Y} \left(1-s_{\{X,Y\}}\bbone\left\{t_A\cap t_B\neq\emptyset\right\}\right)
\]
\[
=-\sum_{A\in X, B\in Y}
\bbone\{t_A\cap t_B\neq\emptyset\}\times
\prod_{{A'\in X, B'\in Y}\atop{(A',B')\neq(A,B)}}
\left(1-s_{\{X,Y\}}\bbone\left\{t_{A'}\cap t_{B'}\neq\emptyset\right\}\right)\ .
\]
Namely, for each `macro' edge $\{X,Y\}$, one needs to choose
an underlying `micro' edge realization $\{A,B\}$, and sum over these choices.
A better rewriting of the outcome is in terms of the
collection $\F\subseteq\PP^{(2)}$ of such edges $\{A,B\}$.
The result is
\[
\prod_{1\le i<j\le p}
\bbone\left\{
[t_{2i-1},t_{2i}]\cap[t_{2j-1},t_{2j}]=\emptyset
\right\}=\sum_{\F}
\int_{[0,1]^{\F}}
\prod_{l\in\F} dv_l
\]
\begin{equation}
\prod_{\{A,B\}\in\F}\left(
-\bbone\{t_A\cap t_B\neq\emptyset\}
\right)
\times
\prod_{\{A,B\}\in\PP^{(2)}\backslash\F}
\left(1-r(\F,v)_{\{A,B\}}\bbone\left\{t_A\cap t_B\neq\emptyset\right\}\right)\ .
\label{BKARoutput}
\end{equation}
The sum is over simple graphs $\F$ on the vertex set $\PP$
which can have at most one edge $\{A,B\}$ with $A\in X$, $B\in Y$,
for any given pair $\{X,Y\}$ of
blocks of the partition $\Pi(\PP+P)$.
No edge $\{A,B\}$ is allowed if it is internal to some block
of the partition $\Pi(\PP+P)$.
Finally, we require that the induced simple graph $\hat{\F}$
on $\Pi(\PP+P)$ be a forest.
The graph $\hat{F}$ is made of all edges $\{X,Y\}\in\Pi(\PP+P)^{(2)}$ 
for which there exists $\{A,B\}\in\F$ with $A\in X$ and $B\in Y$.
If $l=\{A,B\}\in\F$ we use the notation $\hat{l}$
for the edge it thus induces in $\hat{\F}$.
For each $l\in\F$
one associates an integration variable $v_l\in[0,1]$.
Now $\ur(\F,v)$ is the vector
\[
\left(r(\F,v)_{\{A,B\}})\right)_{\{A,B\}\in\PP^{(2)}}
\]
defined as follows.
\begin{itemize}
\item
If $A$, $B$ belong to the same block $X$ of $\Pi(\PP+P)$, then $r(\F,v)_{\{A,B\}}=1$.
\item
If $A$, $B$ belong to different blocks $X$, $Y$ of $\Pi(\PP+P)$ respectively,
then $r(\F,v)_{}=s(\hat{\F},u)_{\{X,Y\}}$ where $u$
is the parameter vector defined by $u_{\hat{l}}=v_l$ for all $l\in\F$.
\end{itemize}

We can now incorporate identity (\ref{BKARoutput}) into our last formula (\ref{ready})
for the partition function:
\[
Z(\alpha,T)=\sum_{p=0}^{\infty} \frac{\alpha^p}{p!}
\sum_{P}
\sum_{\F}
\int_{[0,1]^{\F}}
\prod_{l\in\F} dv_l
\int_{[0,T]^{2p}}
\prod_{i=1}^{2p} dt_i
\ \prod_{j=1}^{p}
\bbone\{t_{2j-1}<t_{2j}\}
\]
\[
\times \prod_{\{A,B\}\in\F}\left(
-\bbone\{t_A\cap t_B\neq\emptyset\}
\right)
\times
\prod_{\{A,B\}\in\PP^{(2)}\backslash\F}
\left(1-r(\F,v)_{\{A,B\}}\bbone\left\{t_A\cap t_B\neq\emptyset\right\}\right)
\]
\begin{equation}
\times
\prod_{\{a,b\}\in\PP} e^{-2|t_a-t_b|}
\times
\prod_{\{a,b\}\in P} h(t_a-t_b)\ .
\label{Zcluster}
\end{equation}
The integrand/summand completely factorizes according to the connected components of the
multigraph $P^{\sharp}+\F$ on $\PP$.
The multigraph $P^{\sharp}$ is made of the edges $\{A,B\}\in\PP^{(2)}$ such that
there exists $a\in A$ and $b\in B$
with $\{a,b\}\in P$.
If two such pairs exist (no more can occur) then the edge $\{A,B\}$ is doubled.
Finally, $P^{\sharp}+\F$
is the superposition of the edges from $P^{\sharp}$ and $\F$.
For instance, after contracting the pairs, a possible configuration is
\[
\parbox{6cm}{
\psfrag{a}{$\scriptstyle{1,2}$}
\psfrag{b}{$\scriptstyle{9,10}$}
\psfrag{c}{$\scriptstyle{7,8}$}
\psfrag{d}{$\scriptstyle{5,6}$}
\psfrag{e}{$\scriptstyle{3,4}$}
\psfrag{f}{$\scriptstyle{11,12}$}
\psfrag{g}{$\scriptstyle{13,14}$}
\psfrag{h}{$\scriptstyle{19,20}$}
\psfrag{i}{$\scriptstyle{17,18}$}
\psfrag{j}{$\scriptstyle{15,16}$}
\includegraphics[width=6cm]{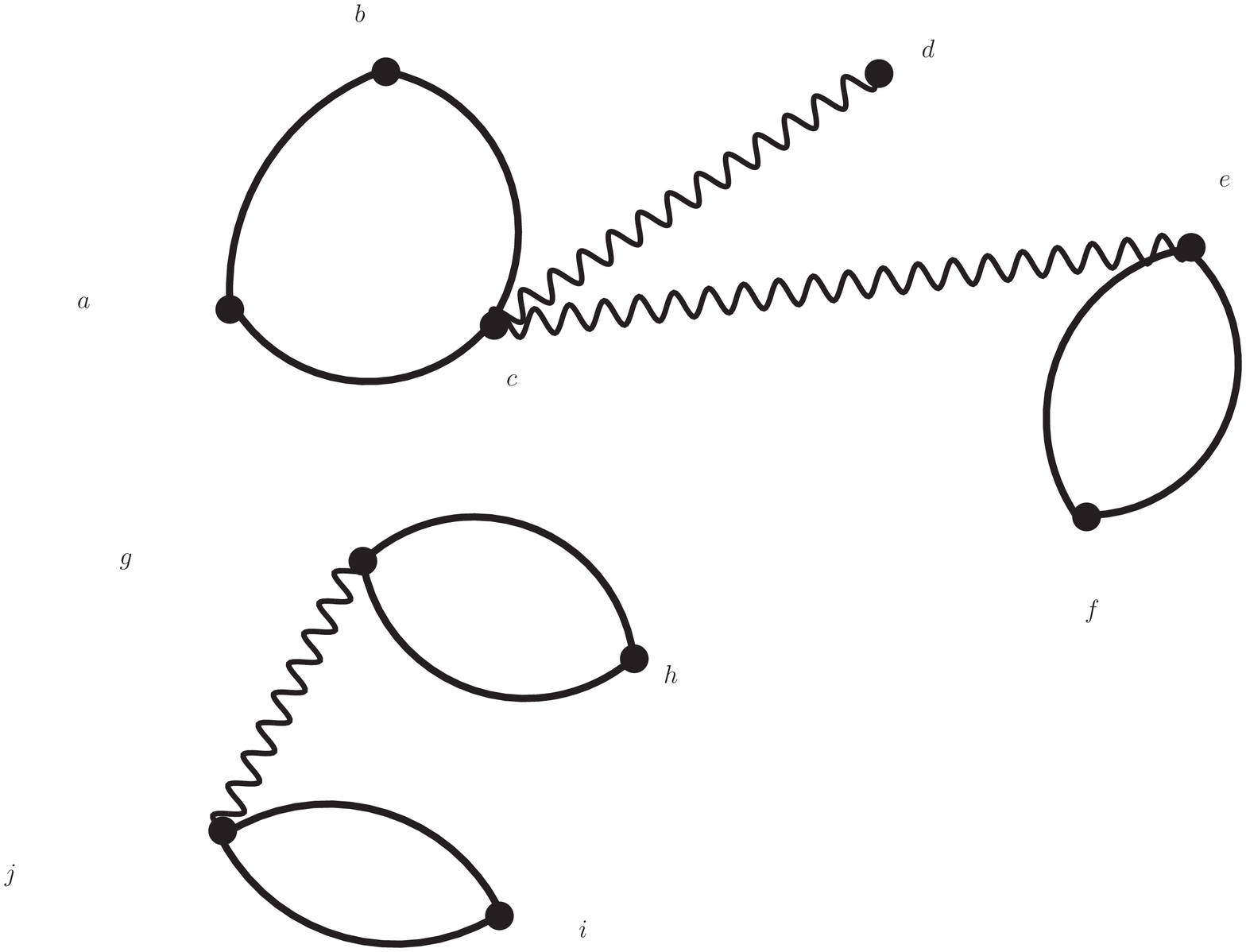}}
\]
where $P^{\sharp}$ is represented by thin lines and $\F$ by squiggly lines.

Let $\Pi\Pi(P^{\sharp}+\F)$
be the partition of $\PP$
made of connected components of $P^{\sharp}+\F$.
The {\it crucial property} of the identity used is that if $A$ and $B$ are not connected
by $P^{\sharp}+\F$, then $r(\F,v)_{\{A,B\}}=0$.
Then complete factorization means, in a first step,
\[
Z(\alpha,T)=\sum_{p=0}^{\infty} \frac{\alpha^p}{p!}
\sum_{P}
\sum_{\F}
\prod_{X\in \Pi\Pi(P^{\sharp}+\F)}
\mathcal{A}(P,\F,X)
\]
where
\[
\mathcal{A}(P,\F,X)=
\int_{[0,1]^{\F_X}}\prod_{l\in\F_X} dv_l
\int_{[0,T]^{2|X|}}
\prod_{i\in\cup_{A\in X} A} dt_i
\]
\[
\prod_{A\in X}
\bbone\{t_{\min A}<t_{\max A}\}
\times
\prod_{\{A,B\}\in\F_X}\left(
-\bbone\{t_A\cap t_B\neq\emptyset\}
\right)
\]
\[
\times
\prod_{\{A,B\}\in X^{(2)}\backslash\F_X}
\left(1-r(\F_X,v_X)_{\{A,B\}}\bbone\left\{t_A\cap t_B\neq\emptyset\right\}\right)
\]
\[
\times
\prod_{\{a,b\}\in X} e^{-2|t_a-t_b|}
\times
\prod_{\{a,b\}\in P_X} h(t_a-t_b)\ .
\]
Here, $\F_X=\F\cap X^{(2)}$ is the restriction of $\F$ to the component $X$.
Likewise, $P_X=P^{\sharp}\cap X^{(2)}$
(as a multiset, i.e., the restriction of the corresponding multiindex in
$\N^{\PP^{(2)}}$ to $X^{(2)}$).
It is easy to see that the computation of $v(\F,v)_{\{A,B\}}$
is purely {\it local} to the component $X$ containing the pair $\{A,B\}$.
Namely, it only involves the knowledge of the data $\F_X$, $P_X$, and $v_X=(v_l)_{l\in\F_X}$.

In a second step, one also has the factorization
\[
\sum_{P}
\sum_{\F}
\prod_{X\in \Pi\Pi(P^{\sharp}+\F)}
\mathcal{A}(P,\F,X)
=\prod_{X\in \Pi\Pi(P^{\sharp}+\F)}
\mathcal{B}(X)
\]
where
\[
\mathcal{B}(X)=
\sum_{P_X}
\sum_{\F_X}
\mathcal{A}(P_X,\F_X,X)
\]
since the combinatorial constraints on $P_X$, $\F_X$
are local to $X$, i.e., do not depend on what happens outside $X$.

Finally, $\mathcal{B}(X)$ only depends on the cardinality of $X$,
i.e., $\mathcal{B}(X)=\mathcal{C}_{|X|}$
where, for any $p\ge 1$,
$\mathcal{C}_p$ is defined as follows. We let
\[
\mathcal{C}_P=\sum_{P}\sum_{\F\ {\rm connecting}}
\int_{[0,1]^{\F}}
\prod_{l\in\F} dv_l
\int_{[0,T]^{2p}}
\prod_{i=1}^{2p} dt_i\ 
\prod_{j=1}^{p}
\bbone\{t_{2j-1}<t_{2j}\}
\]
\[
\times \prod_{\{A,B\}\in\F}\left(
-\bbone\{t_A\cap t_B\neq\emptyset\}
\right)
\times
\prod_{\{A,B\}\in\PP^{(2)}\backslash\F}
\left(1-r(\F,v)_{\{A,B\}}\bbone\left\{t_A\cap t_B\neq\emptyset\right\}\right)
\]
\[
\times
\prod_{\{a,b\}\in\PP} e^{-2|t_a-t_b|}
\times
\prod_{\{a,b\}\in P} h(t_a-t_b)
\]
where the constituents of this formula are defined {\it exactly} as for 
(\ref{Zcluster})
except for one difference:
now we have the extra condition on $\F$ which is that it is 
{\it connecting}, i.e.,
$\Pi\Pi(P^{\sharp}+\F)=\{\PP\}$.
All we just did is use a bijection of $X$ with the set $\PP$ of consecutive pairs
of $[2p]$ with a {\it new} $p$ given by $p=|X|$, in order to {\it relabel} everything.

Now,
\[
Z(\alpha,T)=\sum_{p=0}^{\infty} \frac{\alpha^p}{p!}
\sum_{k\ge 0}
\sum_{{\{X_1,\ldots,X_k\}}\atop{{\rm partition\ of}\ \PP}}
\mathcal{C}_{|X_1|}\cdots \mathcal{C}_{|X_k|}
\]
\[
=\sum_{p=0}^{\infty} \frac{\alpha^p}{p!}
\sum_{k\ge 0}\frac{1}{k!}
\sum_{(X_1,\ldots,X_k)\in\mathcal{P}(\PP)^k}
\bbone\left\{
\begin{array}{c}
{\rm the}\ X_i\ {\rm are\ disjoint}\\
{\rm and}\ \cup X_i=\PP
\end{array}
\right\}
\mathcal{C}_{|X_1|}\cdots \mathcal{C}_{|X_k|}
\]
\[
=\sum_{p=0}^{\infty} \frac{\alpha^p}{p!}
\sum_{k\ge 0}\frac{1}{k!}
\sum_{p_1,\ldots,p_k\ge 1}
\bbone\left\{\sum_{i=1}^{k} p_i=p\right\}
\frac{p!}{p_1!\cdots p_k!}
\mathcal{C}_{p_1}\cdots\mathcal{C}_{p_k}
\]
by the multinomial theorem.

Up to now this was pure algebra and there was no issue of convergence.
Now some analysis is needed in order to write
\[
Z(\alpha,T)=\sum_{k\ge 0}\frac{1}{k!}
\sum_{p_1,\ldots,p_k\ge 1}\ 
\sum_{p\ge 0} \alpha^p\ 
\bbone\left\{\sum_{i=1}^{k} p_i=p\right\}
\ \frac{\mathcal{C}_{p_1}\cdots\mathcal{C}_{p_k}}{p_1!\cdots p_k!}
\]
\[
=\sum_{k\ge 0}\frac{1}{k!}
\sum_{p_1,\ldots,p_k\ge 1}
\left(\frac{\alpha^{p_1}\mathcal{C}_{p_1}}{p_1!}\right)\cdots
\left(\frac{\alpha^{p_k}\mathcal{C}_{p_k}}{p_k!}\right)
=\exp\left(\sum_{p\ge 1}\frac{\alpha^{p}\mathcal{C}_{p}}{p!}\right)\ .
\]
By the the Fubini-Tonelli theorem, all we need in order to justify
these last steps is the absolute convergence
estimate
\[
\sum_{p\ge 1}\frac{|\alpha|^{p}|\mathcal{C}_{p}|}{p!}<\infty
\]
which will follow from (\ref{key}) below.
This granted, we have
\[
\log Z(\alpha,T)=\sum_{p\ge 1}\frac{\alpha^{p}\mathcal{C}_{p}}{p!}\ .
\]
Now use the fact that
\[
\mathcal{C}_p=\int_{[0,T]^{2p}}
\prod_{i=1}^{2p} dt_i\ 
C_p(t_1,\ldots,t_{2p})
\]
with an integrand
\[
C_p(t_1,\ldots,t_{2p})=
\sum_{P}\sum_{\F\ {\rm connecting}}
\int_{[0,1]^{\F}}\prod_{l\in\F} dv_l
\ \prod_{j=1}^{p}
\bbone\{t_{2j-1}<t_{2j}\}
\]
\[
\times\prod_{\{A,B\}\in\F}\left(
-\bbone\{t_A\cap t_B\neq\emptyset\}
\right)
\times
\prod_{\{A,B\}\in\PP^{(2)}\backslash\F}
\left(1-r(\F,v)_{\{A,B\}}\bbone\left\{t_A\cap t_B\neq\emptyset\right\}\right)
\]
\[
\times
\prod_{\{a,b\}\in\PP} e^{-2|t_a-t_b|}
\times
\prod_{\{a,b\}\in P} h(t_a-t_b)
\]
which is defined for any $t$'s in $\R$
and is translation invariant:
\[
C_p(t_1+\tau,\ldots, t_{2p}+\tau)=C(t_1,\ldots,t_{2p}),\ \ \forall \tau .
\]
This allows us to integrate on the more convenient
domain $\left[-\frac{T}{2},\frac{T}{2}\right]^{2p}$,
as well as impose a `pin' $t_1=0$
in the argument of $C_p$.
Hence,
\[
\log Z(\alpha,T)=
\sum_{p\ge 1}\frac{\alpha^{p}}{p!}
\int_{\left[-\frac{T}{2},\frac{T}{2}\right]^{2p}}
\prod_{i=1}^{2p} dt_i
\ C_p(0,t_2-t_1,\ldots,t_{2p}-t_1)
\]
\[
=\sum_{p\ge 1}\frac{\alpha^{p}}{p!}
\int_{-\frac{T}{2}}^{\frac{T}{2}} dt_1
\int_{\R^{2p-1}}
\prod_{i=2}^{2p} ds_i
\ C_p(0,s_2,\ldots,s_{2p})
\]
\[
\times\prod_{i=2}^{2p}
\bbone\left\{
-\frac{T}{2}-t_1\le s_i\le \frac{T}{2}-t_1
\right\}
\]
after changing variables
\[
(t_2-t_1,\ldots,t_{2p}-t_1)\rightarrow (s_2,\ldots,s_{2p})
\]
and implementing the domain specifications by sharp
characteristic functions.
Now change variable to $w=\frac{t_1}{T}$
to get
\[
\frac{\log Z(\alpha,T)}{T}=
\sum_{p\ge 1}\frac{\alpha^{p}}{p!}
\int_{-\frac{1}{2}}^{\frac{1}{2}} dw
\int_{\R^{2p-1}}
\prod_{i=2}^{2p} ds_i\ 
C_p(0,s_2,\ldots,s_{2p})
\]
\[
\times\prod_{i=2}^{2p}
\bbone\left\{
-T\left(\frac{1}{2}+w\right)\le s_i\le T\left(\frac{1}{2}-w\right)
\right\}\ .
\]
For almost every $w$, the characteristic function
goes to $1$ pointwise, as $T\rightarrow\infty$.
By the Lebesgue Dominated Convergence Theorem,
we will
have
\[
\lim_{T\rightarrow\infty} -\frac{\log Z(\alpha,T)}{T}
=
-\sum_{p\ge 1}
\frac{\alpha^p}{p!}
\int_{\R^{2p-1}}
\prod_{i=2}^{2p} dt_i
\ C_p(0,t_2,\ldots,t_{2p})
\]
as soon as one can prove the {\it key estimate}
\begin{equation}
\Gamma_0=\sum_{p\ge 1}
\frac{|\alpha|^p}{p!}
\int_{\R^{2p-1}}
\prod_{i=2}^{2p} dt_i\ 
\left|C_p(0,t_2,\ldots,t_{2p})\right|
<\infty
\label{key}
\end{equation}
which will be done in \S 5.

One easily sees from the previous manipulations that the estimate (\ref{key})
implies the earlier one needed for finite $T$. Namely
one has
\[
\sum_{p\ge 1}\frac{|\alpha|^p |\mathcal{C}_p|}{p!}\le T\cdot \Gamma_0\ .
\]

At last, we can state our main theorem with full precision.

\begin{Theorem}
We have
\[
\lim_{T\rightarrow\infty} -\frac{\log Z(\alpha,T)}{T}
=
-\sum_{p\ge 1}
\frac{\alpha^p}{p!}
\sum_{P}\ \sum_{\F\ {\rm connecting}}
\int_{[0,1]^{\F}}\prod_{l\in\F} dv_l
\int_{\R^{2p-1}}
\prod_{i=2}^{2p} dt_i
\]
\[
\prod_{j=1}^{p}
\bbone\{t_{2j-1}<t_{2j}\}
\times
\prod_{\{A,B\}\in\F}\left(
-\bbone\{t_A\cap t_B\neq\emptyset\}
\right)
\]
\[
\times
\prod_{\{A,B\}\in\PP^{(2)}\backslash\F}
\left(1-r(\F,v)_{\{A,B\}}\bbone\left\{t_A\cap t_B\neq\emptyset\right\}\right)
\]
\[
\times
\prod_{\{a,b\}\in\PP} e^{-2|t_a-t_b|}
\times
\prod_{\{a,b\}\in P} h(t_a-t_b)
\]
where $t_1=0$ by definition.
The limit exists and the sum over $p$
converges
as soon as
$|\alpha|<R_{\rm min}$ with
\[
R_{\rm min}=\left[
32\times \sqrt{e}\times \max\left(||h||_{L^\infty},||h||_{L^1}\right)
\right]^{-1}
\]
which is a lower bound on the analyticity radius.
\end{Theorem}

An example of graph produced by the expansion
(in the contracted pairs representation)
is
\[
\parbox{6cm}{
\psfrag{a}{$\scriptstyle{1,2}$}
\psfrag{b}{$\scriptstyle{9,10}$}
\psfrag{c}{$\scriptstyle{7,8}$}
\psfrag{d}{$\scriptstyle{5,6}$}
\psfrag{e}{$\scriptstyle{3,4}$}
\psfrag{f}{$\scriptstyle{11,12}$}
\psfrag{g}{$\scriptstyle{13,14}$}
\psfrag{h}{$\scriptstyle{19,20}$}
\psfrag{i}{$\scriptstyle{17,18}$}
\psfrag{j}{$\scriptstyle{15,16}$}
\includegraphics[width=6cm]{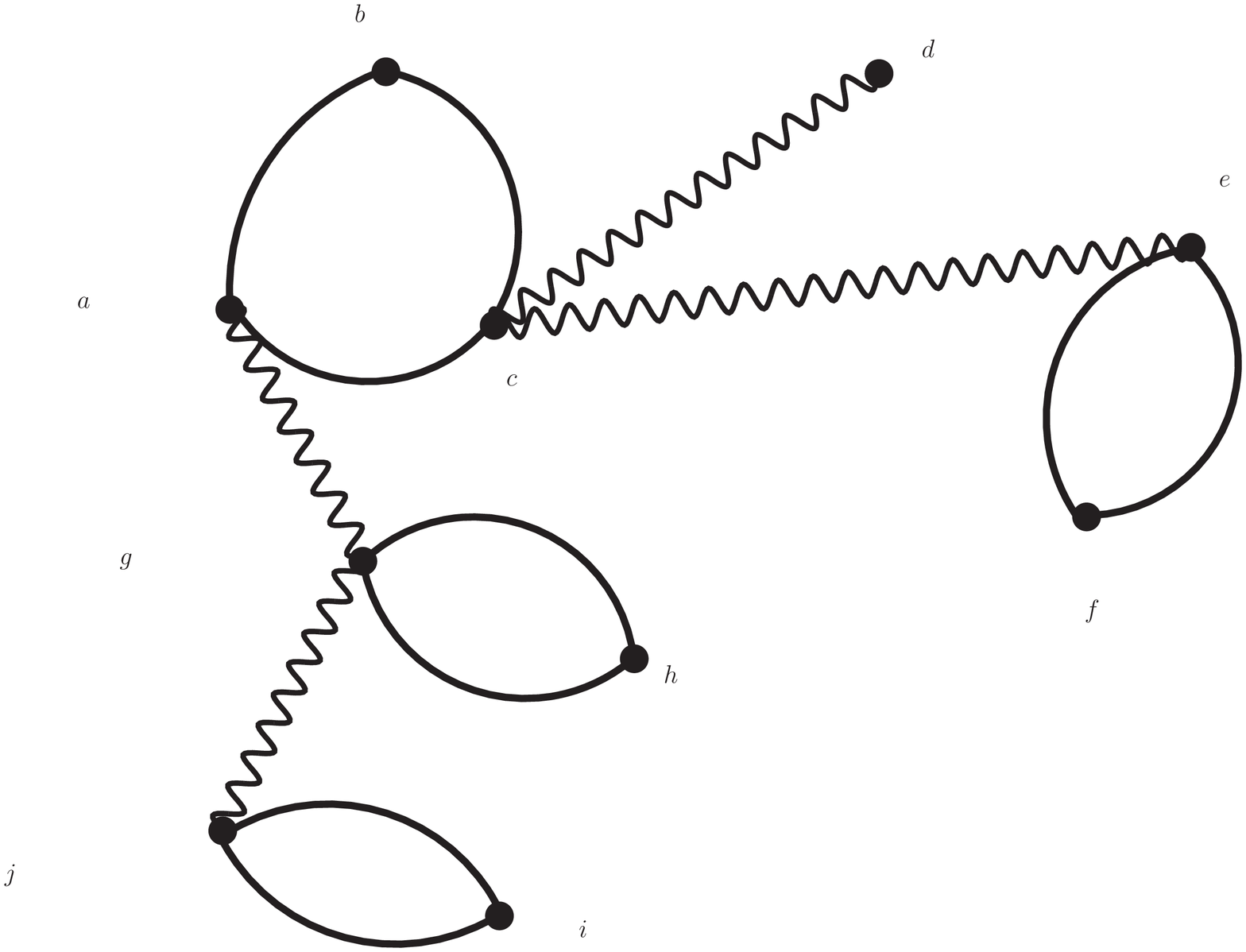}}
\]
where we did not include the interpolated hardcore conditions for a clearer picture.

\section{The key estimate}

We now prove the previous key estimate usually called a polymer bound
in the CQFT literature, or the Koteck\'y-Preiss criterion.
We start from the trivial
bound
\[
\Gamma_0\le
\sum_{p\ge 1}
\frac{|\alpha|^p}{p!}
\sum_{P}\ \sum_{\F\ {\rm connecting}}
\int_{[0,1]^{\F}}\prod_{l\in\F}dv_l
\int_{\R^{2p-1}}
\prod_{i=2}^{2p} dt_i
\]
\[
\prod_{j=1}^{p}
\bbone\{t_{2j-1}<t_{2j}\}
\times\prod_{\{A,B\}\in\F}
\bbone\{t_A\cap t_B\neq\emptyset\}
\]
\[
\times
\prod_{\{A,B\}\in\PP^{(2)}\backslash\F}
\left|1-r(\F,v)_{\{A,B\}}\bbone\left\{t_A\cap t_B\neq\emptyset\right\}\right|
\]
\[
\times
\prod_{\{a,b\}\in\PP} e^{-2|t_a-t_b|}
\times
\prod_{\{a,b\}\in P} h(t_a-t_b)\ .
\]
Since, by definition, the interpolated couplings $r(\F,v)_{\{A,B\}}$
belong to the interval $[0,1]$
we can use
\begin{equation}
\left|1-r(\F,v)_{\{A,B\}}\bbone\left\{t_A\cap t_B\neq\emptyset\right\}\right|\le 1
\label{rbound}
\end{equation}
and also drop the integral over the $v$'s.
Thus,
\[
\Gamma_0\le
\sum_{p\ge 1}
\frac{|\alpha|^p}{p!}
\sum_{P}\ \sum_{\F\ {\rm connecting}}
\mathcal{D}(p,P,\F)
\]
where
\[
\mathcal{D}(p,P,\F)=
\int_{\R^{2p-1}}
\prod_{i=2}^{2p} dt_i
\ \prod_{j=1}^{p}
\bbone\{t_{2j-1}<t_{2j}\}
\times\prod_{\{A,B\}\in\F}
\bbone\{t_a\cap t_B\neq\emptyset\}
\]
\[
\times
\prod_{\{a,b\}\in\PP} e^{-2|t_a-t_b|}
\times
\prod_{\{a,b\}\in P} h(t_a-t_b)\ .
\]
For each cycle of the multigraph $\PP+P$,
delete one edge of $P$.
Let $\PO$ be the graph $P$ with these edges removed.
Clearly,
$|\PO|=|P|-|\pi(\PP+P)|$.
Note that we are simply {\it choosing} an opening of all the cycles.
No summation over these choices is made since we are doing an $L^{\infty}$
rather than an $L^1$ bound.
Now it follows from all the previous combinatorial definitions that $\T=(\PO)^{\sharp}+F$
is a spanning tree which connects the vertex set $\PP$.
Here, $(\PO)^{\sharp}$ is defined in the same way we did for $P^{\sharp}$ earlier.
Also note that $|(\PO)^{\sharp}|=|\PO|$.

We now have
\[
\mathcal{D}(p,P,\F)\le
||h||_{L^\infty}^{|\pi(\PP+P)|}\times
\int_{\R^{2p-1}}
\prod_{i=2}^{2p} dt_i\ 
\prod_{j=1}^{p}
\bbone\{t_{2j-1}<t_{2j}\}
\]
\[
\times
\prod_{\{A,B\}\in\F}
\bbone\{t_A\cap t_B\neq\emptyset\}
\times
\prod_{\{a,b\}\in\PP} e^{-2|t_a-t_b|}
\times
\prod_{\{a,b\}\in \PO} h(t_a-t_b)
\]
and recall that $t_1=0$ by definition.
We now simply use an inductive $L^1$-$L^{\infty}$
bound or `pin and sum' argument for the integration over the $t$'s, by following the tree.
These integrations have to be done in pairs $t_{2i-1},t_{2i}$.
Indeed, what follows should be thought of as a `sum' over all possible {\it intervals}
$t_A$, for $A\in\PP$.
The element $\{1,2\}\in\PP$ is chosen as the root of $\T$ and one bounds the integrals
over the $t_A$'s starting from the leaves, conditioning on their ancestors,
and then progressing towards the root.
The procedure is somewhat subtle since it is a mixture of
straightforward pin and sum with $L^1$ decay
and a generalization of Cammarota's argument~\cite{Cammarota}
to polymers in the continuum instead of on the lattice.
A possible choice of cycle opening for the example in the last picture is
\[
\parbox{6cm}{
\psfrag{a}{$\scriptstyle{1,2}$}
\psfrag{b}{$\scriptstyle{9,10}$}
\psfrag{c}{$\scriptstyle{7,8}$}
\psfrag{d}{$\scriptstyle{5,6}$}
\psfrag{e}{$\scriptstyle{3,4}$}
\psfrag{f}{$\scriptstyle{11,12}$}
\psfrag{g}{$\scriptstyle{13,14}$}
\psfrag{h}{$\scriptstyle{19,20}$}
\psfrag{i}{$\scriptstyle{17,18}$}
\psfrag{j}{$\scriptstyle{15,16}$}
\includegraphics[width=6cm]{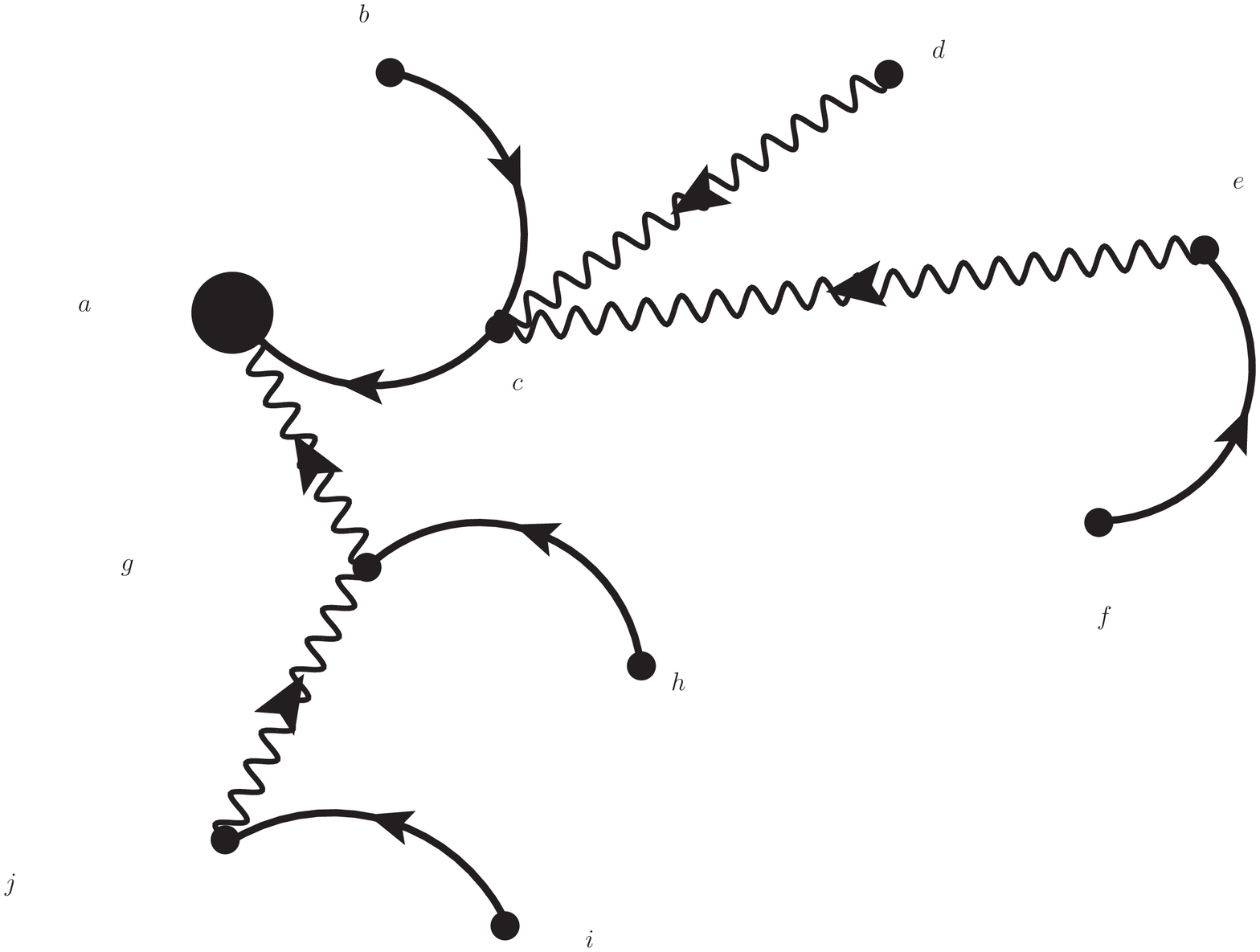}}
\]
where we indicated the orientation towards the root 1,2 for the edges in the remaining tree $\T$.

Let us first explain the integration over $t_{2i-1}$ and $t_{2i}$ when $A=\{2i-1,2i\}$ is
a leaf.

\noindent{\bf 1st case (e.g., pair 17,18 of example):}
If the edge joining $A$ to its ancestor $B$ in the rooted tree $\T$
comes from $(\PO)^{\sharp}$, then one needs to bound an expression of the form
\[
\int_{\R^2}
dt_{2i-1}\ dt_{2i}\ \bbone\{t_{2i-1}<t_{2i}\}
\ e^{-2|t_{2i-1}-t_{2i}|}\ h(t_a-t_b)
\]
where $a\in A$ and $b\in B$.
In this case, first integrate on 
the $t$ labeled by the other element of $A$,
then integrate on $t_a$.
The result is $\frac{1}{2}||h||_{L^1}$.

\noindent{\bf 2nd case (e.g., pair 5,6):}
If the previous edge comes from $\F$ instead,
then one needs to bound
\[
\int_{\R^2}
dt_{2i-1}\ dt_{2i}\ \bbone\{t_{2i-1}<t_{2i}\}
\ e^{-2|t_{2i-1}-t_{2i}|}
\ \bbone\{t_A\cap t_B\neq\emptyset\}\ .
\]
Let $B=\{2j-1,2j\}$ for some $j\neq i$.
Performing the change of variable
\[
\left\{
\begin{array}{ccl}
\xi & = & \frac{1}{\sqrt{2}}(t_{2i}-t_{2i-1})\\
\eta & = & \frac{1}{\sqrt{2}}(t_{2i}+t_{2i-1}-t_{2j}-t_{2j-1})
\end{array}
\right.
\]
the previous integral becomes
\[
\int_{0}^{\infty}
d\xi
\int_{-\xi-\frac{1}{\sqrt{2}}(t_{2j}-t_{2j-1})}^{\xi+
\frac{1}{\sqrt{2}}(t_{2j}-t_{2j-1})}
d\eta\ e^{-2\sqrt{2}\xi}
=\int_{0}^{\infty}
d\xi
\ 2(\xi+\frac{1}{\sqrt{2}}(t_{2j}-t_{2j-1}))\ e^{-2\sqrt{2}\xi}
\]
\[
=\frac{1}{4}\left(1+2(t_{2j}-t_{2j-1})\right)\ .
\]
Therefore, when we later come to doing the integration over $t_B$
we have an extra factor $1+2(t_{2j}-t_{2j-1})$ to control per offspring
born from an $\F$ edge.

\medskip
Let us now consider the integral over $t_A$ where $A$ is a vertex deeper in the tree
but distinct from the root.
Again one has two cases to consider.

\noindent{\bf 1st case (e.g., pair 7,8):}
Same as previous 1st case but with $q$ offsprings connected to $A$
by $(\PO)^{\sharp}$ edges.
One has to bound
\[
\int_{\R^2}
dt_{2i-1}\ dt_{2i}\ \bbone\{t_{2i-1}<t_{2i}\}
\ e^{-2|t_{2i-1}-t_{2i}|}\ h(t_a-t_b)
\ (1+2(t_{2i}-t_{2i-1}))^q\ .
\]
We introduce a parameter $\ga>0$
to be fixed later and we write the trivial inequality
\[
(1+2(t_{2i}-t_{2i-1}))^q\le \frac{q!}{\ga^q}
\cdot
e^{\ga(1+2(t_{2i}-t_{2i-1}))}\ .
\]
Now, provided $\ga<1$,
we can proceed as before and get a bound
\[
\frac{q! e^\ga}{\ga^q}
\int_{\R^2}
dt_{2i-1}\ dt_{2i}\ \bbone\{t_{2i-1}<t_{2i}\}
\ e^{-2(1-\ga)|t_{2i-1}-t_{2i}|}\ h(t_a-t_b)
\]
\[
=\ \frac{q! \ga^{-q} e^{\ga}}{2(1-\ga)}\ ||h||_{L^1}\ .
\]

\noindent{\bf 2nd case (e.g., pair 13,14):}
Same as previous second case but with $q$ offsprings connected to $A$
by $\F$ edges.
One needs to control
\[
\int_{\R^2}
dt_{2i-1}\ dt_{2i}\ \bbone\{t_{2i-1}<t_{2i}\}
\ e^{-2|t_{2i-1}-t_{2i}|}
\ \bbone\{t_A\cap t_B\neq\emptyset\}
\]
\[
\times (1+2(t_{2i}-t_{2i-1}))^q
\]
\[
=\int_{0}^{\infty}d\xi
\int_{-\xi-\frac{1}{\sqrt{2}}(t_{2j}-t_{2j-1})}^{\xi+
\frac{1}{\sqrt{2}}(t_{2j}-t_{2j-1})}
d\eta\ e^{-2\sqrt{2}\xi}
\ (1+2\sqrt{2}\xi)^q
\]
\[
=\int_{0}^{\infty}
d\xi\ e^{-2\sqrt{2}\xi}
\ (1+2\sqrt{2}\xi)^q\times 2(\xi+\frac{1}{\sqrt{2}}(t_{2j}-t_{2j-1}))
\]
\[
\le \int_{0}^{\infty}
d\xi\ e^{-2\sqrt{2}\xi}
\ \ga^{-q} q!\ e^{\ga(1+2\sqrt{2}\xi)}
\times 2(\xi+\frac{1}{\sqrt{2}}(t_{2j}-t_{2j-1}))
\]
\[
=
\frac{e^{\ga} \ga^{-q} q!}{4(1-\ga)^2} 
\left[
1+2(1-\ga)(t_{2j}-t_{2j-1})
\right]\ .
\]
Using $\ga>0$,
we get an upper bound
\[
\frac{q! \ga^{-q} e^\ga}{4(1-\ga)^2} (1+2(t_{2j}-t_{2j-1}))\ .
\]
This is the crux of our proof: one reproduces a factor $1+2(t_{2j}-t_{2j-1})$
of the {\it same form} as that coming from the offsprings.
Therefore, the induction works.

\medskip
Finally,
for the root $A=\{1,2\}$, there is only one case to consider.
Let again $q$ be the number of offsprings attached to it by $\F$ edges.
Given that $t_1=0$, the quantity to bound is
\[
\int_{\R} dt_2\ 
\bbone\{0<t_2\}\ 
e^{-2|t_2|}\ (1+2(t_{2i}-t_{2i-1}))^q
\le \frac{q! \ga^{-q} e^\ga}{2(1-\ga)}\ .
\]

As a result of this discussion, we have
\[
\mathcal{D}(p,P,\F)\le
||h||_{L^\infty}^{|\pi(\PP+P)|}\times
||h||_{L^1}^{|(\PO)^{\sharp}|}
\times
\prod_{A\in\PP}\left[q_A! \ga^{-q_A} \delta(\ga)\right]
\]
where $q_A$ denotes the number of $\F$ offsprings of vertex $A$ and where
\[
\delta(\ga)=\max\left\{
\frac{1}{2}, \frac{1}{4}, \frac{e^\ga}{2(1-\ga)},
\frac{e^\ga}{4(1-\ga)^2}, \frac{e^\ga}{2(1-\ga)}
\right\}\ .
\]
The redundancy recapitulates the previous list of cases.
In fact,
\[
\delta(\ga)=\frac{e^\ga}{4(1-\ga)^2}\times\max\{1, 2(1-\ga)\}
\]
when $0<\ga<1$.
Since
\[
\sum_{A\in\PP} q_A=|F|\le|\T|<p
\]
and
\[
|\pi(\PP+P)|+|\PO|=|P|
\]
we have
\[
\mathcal{D}(p,P,\F)\le
\left[
\max\left(||h||_{L^\infty},||h||_{L^1}\right)
\right]^{|P|}
\times
\delta(\ga)^{|\PP|}
\times \ga^{-p}
\times
\prod_{A\in\PP} q_A!\ .
\]
Since $|P|=|\PP|=p$, we therefore have
\[
\Gamma_0\le
\sum_{p\ge 1}\frac{1}{p!}
\left[
\frac{|\alpha| \delta(\ga)}{\ga}
\max\left(||h||_{L^\infty},||h||_{L^1}\right)
\right]^p
\]
\[
\times
\sum_{P}\sum_{\F\ {\rm connecting}}
\prod_{A\in\PP} q_A!\ .
\]
Now recall that the $q_A$ depend on $p$, $P$ and $\F$.
However, it also depends on the noncanonical choice of cycle opening $\PO$, but
only via the resulting tree $\T$.
Therefore a more precise writing of the right-hand side of the last equation
involves
\[
\sum_{P}\sum_{\F\ {\rm connecting}}
\min_{\T}\left[
\prod_{A\in\PP} q_A(P,\F,\T)!\right]
\]
\[
\le \sum_{P}\sum_{\F\ {\rm connecting}}
\sum_{\T}
\bbone\left\{
\begin{array}{c}
P,\F,\T \\
{\rm are\ compatible}
\end{array}
\right\}
\]
\[
\times
{\rm deg}_{\T}(\{1,2\})!
\prod_{{A\in\PP}\atop{A\neq \{1,2\}}} \left({\rm deg}_{\T}(A)-1\right)!
\]
where compatibility means that $\T$
can be obtained by some choice of cycle opening.
Note the use of $\F\subseteq\T$
to bound the $q$'s in terms of the vertex degrees in $\T$.
Note that the root $\{1,2\}\in\PP$ is special in this regard.
By Fubini,
the bound becomes
\[
\sum_{\T}
{\rm deg}_{\T}(\{1,2\})!
\prod_{{A\in\PP}\atop{A\neq \{1,2\}}} \left({\rm deg}_{\T}(A)-1\right)!
\times
\sum_{P,\F}
\bbone\left\{
\begin{array}{c}
P,\F,\T \\
{\rm are\ compatible}
\end{array}
\right\}
\]
\[
\qquad\le 4^p \sum_{\T}
{\rm deg}_{\T}(\{1,2\})!
\prod_{{A\in\PP}\atop{A\neq \{1,2\}}} \left({\rm deg}_{\T}(A)-1\right)!
\ .
\]
Indeed,
counting the pairs $(P,\F)$
which are compatible with a fixed tree $\T$
can be done by summing
over the number $k=|\pi(\PP+P)|$. This determines the cardinality
of $\F\subseteq\T$, i.e., $|\F|=k-1$.
Finding $\F$ accounts for a factor of $\left(
\begin{array}{c}
p-1\\
k-1
\end{array}
\right)$.
Then $(\PO)^{\sharp}=\T\backslash\F$ is determined.
Since the connected components of $(\PO)^{\sharp}$
are open cycles, there is only one way to close them back, so $P^{\sharp}$ is known.
However, finding $P$ costs an extra $2^p$
due to the choice of micro realizations $\{a,b\}\in [2p]^{(2)}$
of the macro edges $\{A,B\}$ of $P^{\sharp}$.
One gets a $2^p$ bound instead of $4^p$ because $P$ is a perfect matching.
In sum,
\[
\sum_{P,\F}
\bbone\left\{
\begin{array}{c}
P,\F,\T \\
{\rm are\ compatible}
\end{array}
\right\}\le
\sum_{k=1}^{p}
\left(
\begin{array}{c}
p-1\\
k-1
\end{array}
\right) 2^p < 4^p\ .
\]
At last, we can write
\[
\Gamma_0\le
\sum_{p\ge 1}
\frac{1}{p!}
\left[
\frac{4|\alpha|\delta(\ga)}{\ga}
\max\left(||h||_{L^\infty},||h||_{L^1}\right)
\right]^p
\]
\[
\qquad\times
\sum_{\T}
{\rm deg}_{\T}(\{1,2\})!
\prod_{{A\in\PP}\atop{A\neq \{1,2\}}} \left({\rm deg}_{\T}(A)-1\right)!
\]
\[
=\frac{4|\alpha|\delta(\ga)}{\ga}
\max\left(||h||_{L^\infty},||h||_{L^1}\right)
+
\sum_{p\ge 2}
\frac{1}{p!}
\left[
\frac{4|\alpha|\delta(\ga)}{\ga}
\max\left(||h||_{L^\infty},||h||_{L^1}\right)
\right]^p\Bsp
\]
\[
\qquad\qquad
\times
\sum_{{d_1,\ldots,d_p\ge 1}\atop{\Sigma d_i=2p-2}}
\frac{(p-2)!}{(d_1-1)!\cdots(d_p-1)!}\times d_1!(d_2-1)!\cdots (d_p-1)!
\]
by Cayley's theorem~\cite[Theorem 5.3.4]{Stanley}
which counts connecting trees on a labeled set with fixed vertex degrees
$d_i={\rm deg}_{\T}(\{2i-1,2i\})$.

The last sum is bounded using
\[
\sum_{{d_1,\ldots,d_p\ge 1}\atop{\Sigma d_i=2p-2}}
(p-2)!\ d_1\le
(p-1)!
\sum_{{d_1,\ldots,d_p\ge 1}\atop{\Sigma d_i=2p-2}}
1
\]
\[
= (p-1)!
\left(
\begin{array}{c}
2p-3\\
p-1
\end{array}
\right)
\le (p-1)! 4^p\ .
\]
Finally
\[
\Gamma_0\le \sum_{p\ge 1} \frac{(K|\alpha|)^p}{p}
\]
with
\[
K=32\times \max\left(||h||_{L^\infty},||h||_{L^1}\right)
\times
\inf_{0<\ga<1} \left[\frac{\delta(\ga)}{\ga}\right]
\]
\[
\le 32 \times\sqrt{e}\times \max\left(||h||_{L^\infty},||h||_{L^1}\right)
\]
if one takes $\ga=\frac{1}{2}$ for simplicity.
\qed

As an illustration of the quality of these bounds, let us pick
$f(k)=\bbone\{|k|\le 1\}$
and $\omega(k)=|k|$.
In this case
\[
||h||_{L^\infty}=||\frac{f}{\sqrt{\omega}}||_{L^2}=4\pi\int_0^1 k dk=2\pi
\]
and
\[
||h||_{L^1}=8\pi \int_{0}^{\infty}ds \int_{0}^{1} dk\ k\ e^{-sk}=8\pi 
\]
and therefore we have a lower bound on the radius of analyticity in $\alpha$
\[
(256\pi \sqrt{e})^{-1}\simeq 7.54\times 10^{-4}\ .
\]
Since $\alpha=\left(\frac{\lambda}{4\pi}\right)^2$,
this translates (provided Bloch's formula holds)
into
a lower bound 
for the $\lambda$ radius of analyticity $\simeq 0.34$.

\section{Closing remarks}

\noindent{\bf Remark 1:}
One could presumably improve the estimates, in the spirit
of~\cite{FernandezP}, using the remaining hardcore constraints which were thrown away in (\ref{rbound}).
In this case one should use the Rooted Taylor Forest Formula of~\cite{AbdesselamR}
which forces offspring exclusion. Then one should follow the presentation in~\cite{Faris}.
The rooted forest formula has more of a Fermionic flavor. It is related to the matrix-tree theorem
and is closer to Penrose's lemma for the convergence of the Mayer series~\cite{Penrose}.

\medskip
\noindent{\bf Remark 2:}
Surprisingly, in two dimensions, the obtained series still makes sense, order-by-order, even without
infrared cut-off. Its Borel transform has a nonzero radius of analyticity.
This can be shown by a more refined analysis along the lines of the previous proof.
In this case $h(s)\sim\frac{1}{s}$ for large $s$.
Therefore, one cannot open the cycles. The integral over the $t$'s belonging to such a cycle
of length $k$ (the number of $\PP$ points involved) converges but produces a $k!$ (see,
e.g.,~\cite[\S 3]{ProcacciPNM}).
In two dimensions, $\inf \sigma(H_\lambda)=-\infty$ and we do not know what the term-by-term
infrared-finite perturbation series means in that case.

\medskip
\noindent{\bf Remark 3:}
There are similarities between our expansion and the one found, e.g., in~\cite{DeRoeckK},
and for a good reason: both essentially are resummations of the same underlying
Feynman diagram expansion. Indeed, the massless spin-Boson model
is a typical example for which
the results of~\cite{DeRoeckK} can be applied.
There are also important differences. The article~\cite{DeRoeckK} uses a real-time setting
whereas our article is framed in the Euclidean setting.
Also, the cited article uses a decomposition into cubes or rather
intervals as is traditional in Bosonic CQFT.
Our approach does not use such a decomposition and is closer to the Feynman diagram
perturbation series. We decouple abstract vertices of such diagrams rather than
cells in a decomposition of space-time. 
A similar distinction can be made between existing techniques for the treatment
of Fermionic theories. For instance, early cluster expansions for Fermions
used decompositions into cells (see, e.g.,~\cite{GawedzkiK,FeldmanMRS}).
Later, expansions which decouple Feynman diagram
vertices~\cite{Lesniewski,AbdesselamR2} allowed to achieve the same convergence bounds
in a simpler way.

\medskip
\noindent{\bf Remark 4:}
The combinatorial structures which feature in our expansion for the spin-Boson model are
more than strikingly similar to the ``canvases'' of~\cite{AbdesselamB} which are related to the
loop-erased random walk (LERW). We suspect
a model mapping such as the one in~\cite{BrydgesI} is lurking behind this similarity.
Here the LERW seems to be one on the real line where a particle alternately `marks its territory'
with an interval of exponentially distributed length, and diffuses according to the function $h$.
The self-avoidance (or rather loop-erasure prescription) is in terms of this previously 
marked territory.
We believe a similar LERW model on $\R^2$ instead of $\R$ could help bridge
the gap between the Schramm-L\"owner Evolution theory~\cite{Schramm} and conformal field
theory in its Coulomb gas formulation (see, e.g.,~\cite{DiFrancescoMS}).

\bigskip
\noindent{\bf Acknowledgements:}
{\small
This work was supported in part by the National Science Foundation 
under grant DMS \# 0907198.
We are indebted to Ira Herbst for introducing us to the spin-Boson model and for many helpful
discussions. We also thank Lawrence Thomas for the suggestion of studying
the ground state
energy of the spin-Boson model via the Ising model on the real line.
Finally we thank David Hasler and Wojciech De Roeck for insightful discussions.}

\end{document}